\documentclass[11pt]{article} 

\usepackage[utf8]{inputenc} 
\usepackage{authblk}

\usepackage{color}

\usepackage{geometry} 
\usepackage{graphicx} 
\usepackage{cite}
\usepackage{amsmath}
\usepackage{booktabs} 
\usepackage{array} 
\usepackage{paralist} 
\usepackage{verbatim} 
\usepackage{subfig} 

\usepackage{fancyhdr} 
\usepackage{sectsty}
\allsectionsfont{\sffamily\mdseries\upshape} 

\usepackage[nottoc,notlof,notlot]{tocbibind} 
\usepackage[titles,subfigure]{tocloft} 

\title{Fluctuating hydrodynamics and mesoscopic effects of spatial correlations in dissipative systems with conserved momentum}
\author[1]{A. Lasanta}
\author[1]{A. Manacorda}
\author[2]{A. Prados}
\author[1]{A. Puglisi}
\affil[1]{Istituto dei Sistemi Complessi - CNR and Dipartimento di Fisica, Universit\`a di Roma Sapienza, P.le Aldo Moro 2, 00185, Rome, Italy}
\affil[2]{F\'{\i}sica Te\'orica, Universidad de Sevilla, Apartado  de Correos 1065, E-41080 Sevilla, Spain}

\begin{document}
\maketitle
\begin{abstract}
  We introduce a model described in terms of a scalar velocity field
  on a 1d lattice, evolving through collisions that conserve momentum
  but do not conserve energy. Such a system posseses some of the
    main ingredients of fluidized granular media and naturally
    models them.  We deduce non-linear fluctuating hydrodynamics
  equations for the macroscopic velocity and temperature fields, which
  replicate the hydrodynamics of shear modes in a granular
  fluid. Moreover, this Landau-like fluctuating hydrodynamics predicts
  an essential part of the peculiar behaviour of granular fluids, like
  the instability of homogeneous cooling state at large size or
  inelasticity. We compute also the exact shape of long range spatial
  correlations which, even far from the instability, have the physical
  consequence of noticeably modifying the cooling rate. This
    effect, which stems from momentum conservation, has not been
    previously reported in the realm of granular fluids.
\end{abstract}

\section{Introduction}

Since the seminal paper of Einstein \cite{eins}, it is well known that
the fluctuating behaviour of systems at the mesoscopic level reflects
the hectic microscopic dynamics beneath. While the equilibrium
behavior of mesoscopic fluctuations has been investigated and
understood in detail~\cite{onsager,landau}, a big effort is still
being carried out to explore the fluctuating properties of
non-equilibrium media~\cite{Bertini}. These are known to lead, in
great generality, to the emergence of spatial correlations and pattern
self-organization~\cite{grinstein,correlations}.  In this context, the
crucial task of connecting microscopic and mesoscopic dynamics is
considerably simplified when {there exists a} separation of scales,
which makes it possible to introduce {\em slow} fields evolving
under the so-called hydrodynamic equations~\cite{kl}.

An important class of systems exhibiting patterns includes two types
of complex fluids: active matter~\cite{active,yeomans}, such as bacteria or
birds, and fluidized granular
materials~\cite{jaeger96}. Interestingly, active and granular matter
are often associated~\cite{activegranular,baskaran,NRyM07,ASOyU08 }. They are not only
relevant for applied and biomedical sciences, but also offer
fascinating challenges for kinetic theory~\cite{puglbook}. Indeed, the
lack of energy conservation in the microscopic dynamics makes them
intrinsically out-of-equilibrium systems~\cite{goldhirsch}. In
granular and active fluids, the spectacular emergence of spatial
patterns, particularly in vectorial fields such as momentum or
orientation, is often understood in terms of hydrodynamic
equations~\cite{active2}.  Furthermore, a relevant role is played by
fluctuations, as an inevitable consequence of the relative small
number of their elementary constituents~\cite{cgm}.

One of the most intensively investigated states in the realm of
  granular fluids is the homogeneous cooling state (HCS)
  \cite{PyL01,Go03}. Therein, the granular temperature decays in time
  following Haff's law \cite{haff}, whereas the system remains
  spatially homogeneous. Remarkably, the HCS is the reference state
  for the hydrodynamic description of granular fluids but is unstable:
  for large enough inelasticity or system size, the scaled
  fluctuations of the transverse velocity increase (shear instability)
  and eventually density inhomogeneities arise (clustering
  instability) \cite{GyZ93}. This instability for large system sizes
  makes it relevant to look into the finite size corrections to the
  physical quantities, like the cooling rate, since a ``thermodynamic
  limit'' in which the system size is infinitely large cannot be taken
  without simultaneously making the inelasticity
  vanish. Notwithstanding, and to the best of our knowledge, these
  finite size corrections have only been investigated for a system of
  smooth inelastic hard spheres very close to the shear instability
  \cite{BDGyM06}.

In deriving mesoscopic transport equations from microscopic rules,
analytical results are needed and simple models are good candidates
for this~\cite{kmp,active3}.  In this paper, we study a 1d lattice
model which implements two main ingredients of granular fluids:
inelastic collisions and momentum conservation. Given the
  simplicity and appealing physical picture of the model, this novel
  approach may help to improve our current understanding of the
  complex behaviour of granular fluids. To start with, we recover some
  of the main features of granular fluids: more specifically, the
  large size and inelasticity shear instability. In addition, we are
  able to compute exactly the shape of velocity correlations, which
  allows us to, first, extend the known results of fluctuating
  hydrodynamics ~\cite{Noijeernst} and, second, obtain the finite size
  corrections to the cooling rate. The latter has a non-trivial
  dependence on the inelasticity and system size: for a given
  inelasticity, it changes sign at a certain value of the system size
  that is smaller than the one corresponding to the shear
  instability.

Finally, we would like to stress that momentum conservation is a
physical constraint that certainly has a recognised role in the
appearance of long range spatial
correlations~\cite{correlations,rama}. However, since it
complicates the description and the derivation of exact results,
it is rarely considered in its entirety. Here, starting from the
microscopic dynamics, we are able to rigorously derive the mesoscopic
equations that describe the average and fluctuating behaviour at the
hydrodinamic scale, taking into account momentum conservation in full.

\section{Microscopic equations of the model.}
Fluctuating hydrodinamics in linear and nonlinear lattice diffusive
models {have been extensively studied in recent years, both} in
the conservative \cite{Pablo,Pablo2,Pablo3,HyK11} and in the
dissipative cases \cite{PLyH12a,PLyH11a,PLyH13}.  Inspired {by}~\cite{Balbet}, {we consider a} 1d
lattice with $N$ sites {and} given boundary conditions,
{either} periodic or thermostatted, depending on the situation
of interest. At a given time $p$, each site $l$ possesses a velocity
$v_{l,p}$ and the total energy of the system {is}
$E_{p}= \sum_{l=1}^{N} v_{l,p}^{2}$. In an elementary time step of the
dynamics, with a probability discussed below, a pair of nearest
neighbors $\left( l,l+1 \right)$ collides inelastically and evolves
following the rule ($0<\alpha\leq 1$)
\begin{equation}\label{coll}
v_{l,p+1} = v_{l,p}-(1+\alpha)\Delta_{l,p}/2, \qquad
v_{l+1,p+1} = v_{l+1,p}+(1+\alpha)\Delta_{l,p}/2,
\end{equation}
 having defined the relative velocity
\begin{equation}
\Delta_{l,p}=v_{l,p}-v_{l+1,p}.
\end{equation} 
Momentum is always conserved,
$v_{l,p}+v_{l+1,p}=v_{l,p+1}+v_{l+1,p+1}$, while energy, if
$\alpha \neq 1$, is not:
$v^{2}_{l,p+1}+v^{2}_{l+1,p+1}-v^{2}_{l,p}-v^{2}_{l+1,p}=
(\alpha^2-1)\Delta_{l,p}^{2}/2<0$.

\begin{figure}
\centering
  \includegraphics[angle=90,width=0.78\textwidth]{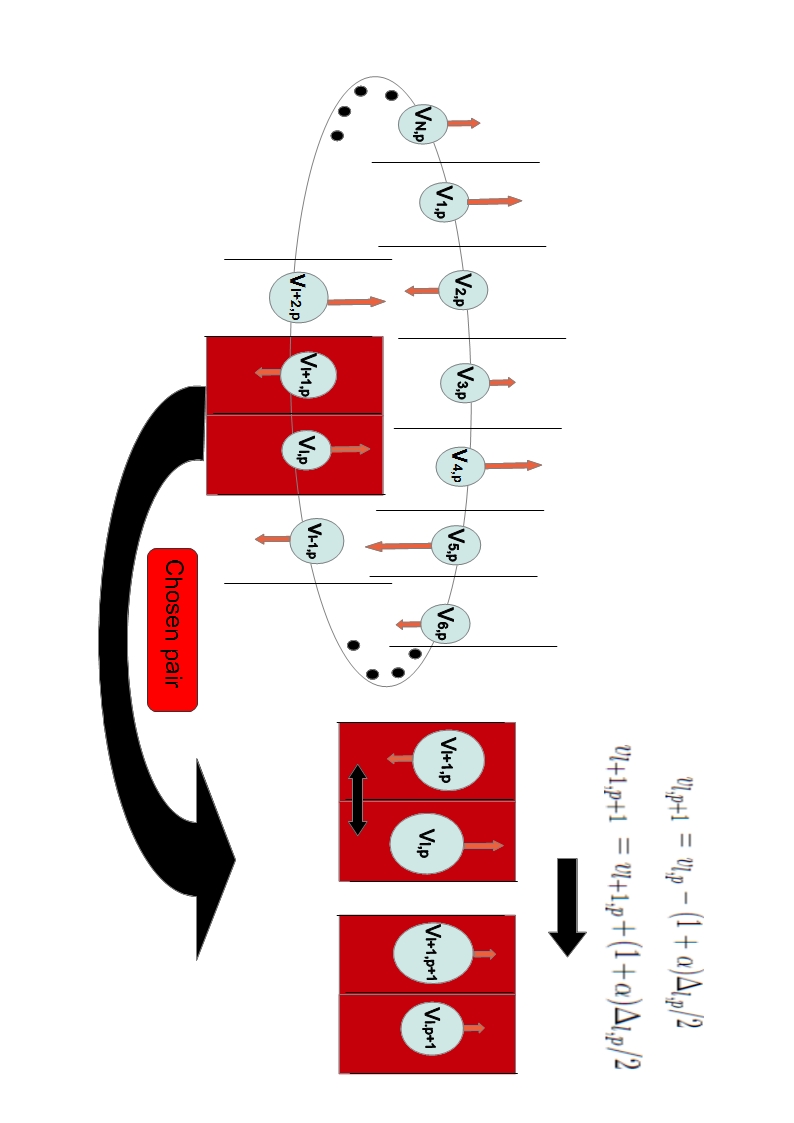}
  \caption{The model is defined on a lattice, each site being
    characterized by a velocity $v_{l}$ and standing for a fluid
      slab. The dynamics proceeds via nearest-neighbour collisions as
    defined in the text, in which part of the energy is
    dissipated. This model resembles a sheared granular system
    at the mesoscopic level. To be specific, we show the sketch that
    corresponds to periodic boundary conditions.}
\label{sketch}
\end{figure}

The definition of the model implies that there is no mass transport,
particles are at fixed positions and they only exchange momentum and
kinetic energy. We are also disregarding the so-called kinematic
constraint in~\cite{Balbet}, namely a colliding pair is chosen
independently of the sign of its relative velocity. {This can be
  understood as} the velocity of the particles {representing not}
their motion along the lattice axis but rather {along} a transversal
one: {in fact,} the hydrodynamics derived here replicates transport
equations for granular gases in $d>1$ restricted to the shear
(transverse) velocity field, see Fig.~\ref{sketch}. 

In the context of granular fluids, the model may be physically
  motivated as follows. We start from a $d>1$ system that has been divided into
  ``slabs'' that are perpendicular to the lattice
  direction. Specifically, each particle on the lattice represents one
  slab. In this sense, the parameter $\alpha$ that appears in the
  collision rule \eqref{coll} should not be confused with the usual
  restitution coefficient defined in granular media, since here
  $\alpha$ stands for an \textit{effective} inelasticity for the
  collisions between slabs. The connection with a ``real'' granular
  fluid should be done at the level of the cooling rate
  that appears in the hydrodynamic equations, see Section
  \ref{sec:hydro}. 

Now we write down the evolution equation for the velocities.  At time
$p$, the probability that the nearest-neighbours {pair} at
{sites} $(l,l+1)$ collide {is assumed to be}
$P_{l,p} \propto |\Delta_{l,p}|^{\beta}$. {Then,}
\begin{equation}\label{3.4}
v_{l,p+1}-v_{l,p}=-j_{l,p}+j_{l-1,p}, \qquad j_{l,p}=(1+\alpha)\Delta_{l,p}\delta_{y_p,l}/2,
\end{equation}
which is a discrete continuity equation for the (conserved) momentum,
with $j_{l,p}$ being the momentum current, that is, the flux of
momentum from site $l$ to site $l+1$ at the time step $p$.  Therein,
$\delta_{y_{p},l}$ is Kronecker's $\delta$ and $y_{p}$ is a random
integer which selects the colliding pair {with probability $P_{l,p}$}.
The evolution equation for the energy is obtained by squaring
(\ref{3.4}),
\begin{eqnarray}\label{3.7}
v^2_{l,p+1}-v^2_{l,p}=-J_{l,p}+J_{l-1,p}+d_{l,p}.
\end{eqnarray}
Again, we have defined an energy current $J_{l,p}$ from site $l$ to
site $l+1$ and the energy dissipation $d_{l,p}$ at site $l$
as 
\begin{equation}
J_{l,p}=(v_{l,p}+v_{l+1,p})j_{l,p}, \qquad d_{l,p}= \frac{\alpha^2-1}{4}(\delta_{y_p,l}\Delta_{l,p}^2
+\delta_{y_p,l-1}\Delta_{l-1,p}^2 )<0.
\end{equation}
The sink term $d_{l,p}$ only vanishes in the elastic case $\alpha=1$.

Stochastic simulations of the model are done as follows. At any
  Montecarlo step $p$, one site $l$ is picked with probability
  $P_{l,p}\propto |\Delta_{l,p}|^{\beta}$ and particles $l$ and $l+1$ collide following the
  microscopic rules (\ref {coll}). The simplest choice for $P_{l,p}$
  corresponds to $\beta=0$, all the pairs are chosen with uniform
  probability, $P_{l,p}=1/L$, in which $L$ is the number of
  pairs. This is often called in the literature the model of inelastic
  Maxwell molecules (MM)~\cite{Balbet}.  Note that $L$ is basically
  equal to $N$ but depends on the boundary conditions: for periodic
  boundary conditions, it is $L=N$, but if we consider the system
  coupled to two extra sites $0$ and $N+1$, which introduce the
  appropriate boundary conditions, it is $L=N+1$. The periodic
  boundary conditions, sketched in figure \ref{sketch}, correspond to
  the free (undriven) evolution of the system and if $l=N$ it is the
  pair $(N,1)$ that collides.

In the following, we discuss {the} hydrodynamic limit, fluctuations
and correlations for a particular choice of $P_{l,p}$. Specifically,
we consider the case of MM: such a choice is dictated by the
will of simplifying the presentation and making clear the essential
points.  We postpone a more complete and general discussion to a more
technical and detailed paper, in preparation. The theoretical
  results are compared to the numerical simulations described above. A
  large enough value of $L$, which is indicated in the figures, has
  been used to ensure the hydrodynamic limit, and we have averaged
  over $10^5$ realizations of the stochastic dynamics. Aside from MM,
  results for HS ($\beta=1$) are also shown in a few, clearly marked,
  cases.

\section{\label{sec:hydro} Hydrodynamic limit: average equations and fluctuations}

Let us define, as usual, the following local averages over initial
conditions and noise realizations:
$u_{l,p}\equiv \langle v_{l,p}\rangle$,
$E_{l,p} \equiv \langle v_{l,p}^{2}\rangle$ and
$T_{l,p}\equiv E_{l,p}-u_{l,p}^{2}$. Their evolution is obtained under
a series of assumptions.  With the choice of MM, $y_{p}$ is an uniform
distributed random integer, namely $\langle\delta_{y_p,l}\rangle=1/L$,
with $L$ being the number of nearest neighbor pairs.  In addition,
when considering the average dissipation at site $l$, there appears
moments like $\langle v_{l,p}v_{l\pm 1,p}\rangle$.
To the lowest order, we assume that neighbouring velocities are
  uncorrelated, that is, 
$\langle v_{l,p}v_{l\pm 1,p}\rangle=u_{l,p}u_{l\pm
  1,p}$ (see Appendix A).

Now we assume that $u_{l,p}$ and $E_{l,p}$ are smooth functions of
space and time and introduce a continuum (``hydrodynamic'') limit (HL)
by defining macroscopic scales: $\Delta x \sim L^{-1}$ and
$\Delta t \sim L^{-3}$. Each spatial derivative introduces thus a
factor $L^{-1}$ in the continuum limit: therefore, the difference between
the current terms in the balance equations is of the order of $L^{-3}$.
On the other hand, the dissipation goes as $(1-\alpha^{2})L^{-1}$, which
makes it useful to define the cooling rate as (see Appendix \ref{appa})
\begin{equation}
  \label{eq:nu}
  \nu=(1-\alpha^{2})L^{2}.
\end{equation}
It is natural, on the scales defined by the HL, to define the
mesoscopic fields $u(x,t)$, $E(x,t)$ and $T(x,t)$, as well as the
average mesoscopic currents
$j(x,t)=\lim_{L\to\infty}L^{2}\langle j_{l,p}\rangle=-\partial_{x}
u(x,t)$,
$J(x,t)=\lim_{L\to\infty}L^{2}\langle
J_{l,p}\rangle=-\partial_{x}\left[ u^{2}(x,t)+T(x,t)\right]$
and the average mesoscopic dissipation of energy
$d(x,t)=\lim_{L\to\infty}L^{3}\langle d_{l,p}\rangle=-\nu T(x,t)$,
which depends only on the granular temperature but not on the average
local velocity $u(x,t)$, as expected on a physical basis. After
computations detailed in Appendix \ref{appa}, we
get the HL of~\eqref{3.4} and~\eqref{3.7}, which are
$\partial_t u(x,t) = -\partial_x j(x,t)$ and
$\partial_t E(x,t)=-\partial_x J(x,t) -\nu T(x,t)$ respectively. Then,
we can write the average hydrodynamic evolution equations
\begin{equation}\label{3.16}
\partial_{t} u=\partial_{xx}u, \qquad
\partial_{t}T=-\nu T+ \partial_{xx}T+2\left(\partial_{x}u\right)^2 .
\end{equation}
over the length and time scales defined above. 
Note that,
here, we have substituted $1+\alpha$ by $2$, because
$\alpha^{2}=1-\nu L^{-2}$, and we have already neglected $L^{-1}$
terms. These equations must be supplemented by appropriate boundary
conditions for the {situation} of interest. Technical details
are deferred to a later paper. 

Let us consider the fluctuations of the microscopic currents and
dissipation, that is, $j_{l,p} = \widetilde{j}_{l,p} + \xi_{l,p}$,
$J_{l,p}=\tilde{J}_{l,p}+\eta_{l,p}$, and
$d_{l,p}=\tilde{d}_{l,p}+\theta_{l,p}$. Tilde variables correspond to
a partial average: they are averaged over the fast variables $y_{l,p}$
but not over the slow ones $v_{l,p}$. Thus, for example,
$\widetilde{j}_{l,p}=(1+\alpha)\Delta_{l,p}/2L$. It is clear that this
choice guarantees that all noises $\xi_{l,p}$, $\eta_{l,p}$ and
$\theta_{l,p}$ have zero average.  The noise correlations read
$\langle \xi\xi' \rangle \sim A_{\xi}\delta(x-x')\delta(t-t')$,
$\langle \eta\eta' \rangle \sim A_{\eta} \delta(x-x')\delta(t-t')$ and
$\langle\theta\theta'\rangle\sim A_{\theta}\delta(x-x') \delta(t-t')$
with amplitudes $A_{\xi}=2L^{-1}T(x,t)$,
$A_{\eta}=4L^{-1}T(x,t)[T(x,t)+2u^2(x,t)]$ and
$A_{\theta} =3L^{-3} \nu^2 T^2(x,t)$ (see Appendix \ref{appb}). {In
  the above relations, we have used the notation $\xi\equiv\xi(x,t)$
  and $\xi'\equiv \xi(x',t')$, and similar notations for
  $\eta,\eta',\theta,\theta'$.}  Thus, the currents noises are delta
correlated in space and time, and their amplitudes scale as $L^{-1}$
with the system size $L$.  On the other hand, the noise of the
dissipation is subdominant with respect to the moment and energy
currents, its amplitude being proportional to $L^{-3}$, and therefore
it is usually negligible. Gaussianity for these noises can be easily
demostrated, see \cite{PLyH12a}. Interestingly, being in the presence
of two fluctuating fields, correlations between different noises
appear.  Theoretical predictions for noise correlations, amplitudes
and Gaussianity have been succesfully tested in both MM and HS
simulations, see Appendices ~\ref{appb} and \ref{appc}.

\section{Solutions, HCS and instabilities}

Here we focus our attention on the case of spatial periodic boundary
conditions and an initial ``thermal condition'': $v_{l,0}$ is a random
Gaussian variable with zero average and unit variance, that is,
$T_{l,0}\equiv T(x,0)=1$. Starting from this condition, the system
is expected to typically fall into the so-called Homogeneus
Cooling State (HCS), in which the velocity and temperature
fields remain spatially uniform, and the temperature decays
in time. Indeed, the solution of the average hydrodynamic
equations~(\ref{3.16}) reads
\begin{equation}
u(x,t) = 0, \qquad
T_{HCS}(x,t) = T(t=0) e^{-\nu t} .
\label{HCS}
\end{equation}
The exponential decrease of the granular temperature is typical of MM,
where the collision frequency is velocity-independent. It
replaces the so-called Haff's law which was originally derived in the
HS case, where $T_{HCS}\sim t^{-2}$ because
$\dot{T} \propto T^{3/2}$~\cite{haff}.

\begin{figure}
\centering
\includegraphics[angle=0,width=0.6\textwidth]{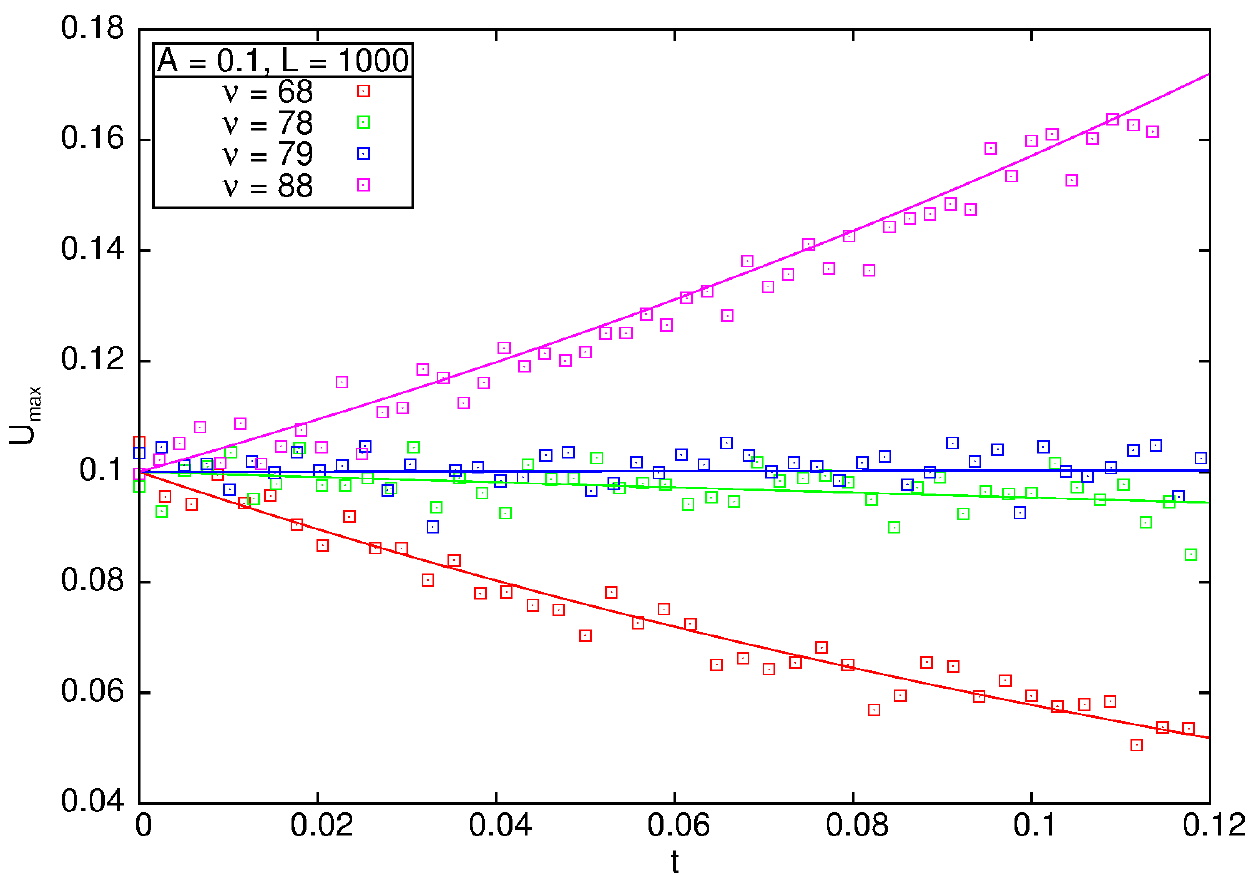}
\caption{\small Rescaled velocity profile maximum $U_{max}=U(x_M,t)$
  as a function of time, where $x_M = 1/4$.   Trajectories start from a sinusoidal average velocity profile
  $u(x,0)=u_0 \sin(2 \pi x)$ (here $u_0=0.1$), which gives
  hydrodynamic predictions  $U(x,t) = u_0 \sin(2 \pi x) e^{(\nu-\nu_c)t/2}$ (drawn as solid lines).  }
\label{f:uinstability}
\end{figure}

The HCS is known to be unstable: it breaks down in too large or too
inelastic systems \cite{MN93}. In our model and in the
hydrodynamic limit, this condition is expected to be replaced by a
condition of large $\nu$. The stability is studied by introducing
rescaled fields $U(x,t)=u(x,t)/\sqrt{T_{HCS}(t)}$ and
$\tilde{T}=T(x,t)/T_{HCS}(t)$ and by linearizing the hydrodinamic
equations near the HCS, i.e. $T(x,t)=T_{HCS}(t)+\delta T(x,t)$ and
$U(x,t)=\delta U(x,t)$. The analysis of linear equations becomes
straightforward by space-Fourier-transforming, which gives
\begin{equation}\label{3.1.4}
\partial_{t}\delta U(k,t)=\frac{\nu-2k^{2}}{2}\delta U(k,t),\qquad
\partial_{t}\delta \tilde{T}(k,t)=-k^2\delta \tilde{T}(k,t).
\end{equation}
Therefore, $\delta U$ is unstable for wave numbers that verify
$\nu-2k^2>0$. In the continuous variables we are using, the system
size is $1$, so that the minimum available wavenumber is
$k_{min}=2\pi$.  Thus, {there} is no unstable {mode} for
{$\nu$ (lengths) below a certain threshold $\nu_{c}$ ($L_{c}$), with}
\begin{equation}
  \label{eq:2}
\nu_{c}=8\pi^{2}, \qquad   L_{c}=2\pi\sqrt{2}\left(1-\alpha^{2}\right)^{-1/2} .
\end{equation}
On the contrary, for $\nu>\nu_c$ {($L>L_{c}$)}, the HCS is
unstable and modes with wave numbers verifying $k<\sqrt{\nu/2}$
increase with time. This instability mechanism is identical to the
one found in granular gases for shear modes~\cite{Noijeernst}.
Theoretical predictions and simulations results perfectly agree, as plotted in
Fig.~\ref{f:uinstability}. It is important to stress that the
amplification appears in the rescaled velocity $U(x,t)$ and not in the
velocity $u(x,t)$. The same result is found and  compares well with simulations in the HS case.

\section{\label{sec:corr}Spatial correlations and their effects in the HCS}
Assuming space
translation invariance, which is certainly valid in the HCS, we can
write a hierarchy of equations for the spatial correlations defined as
$C_{k,p}=\langle v_{j,p}v_{j+k,p}\rangle$ at a distance of $k$ sites,
at time $p$:
\begin{eqnarray}
C_{0,p+1} & = & C_{0,p}+(\alpha^2-1)L^{-1}(C_{0,p}-C_{1,p}), \label{hier} \\
C_{1,p+1} & = & C_{1,p}+(1-\alpha^2)L^{-1}(C_{0,p}-C_{1,p})+
    (1+\alpha)L^{-1}(C_{2,p}-C_{1,p}), \label{hier2} \\
C_{k,p+1} & = &C_{k,p}+(1+\alpha)L^{-1}(C_{k+1,p}+C_{k-1,p}-2C_{k,p}), \quad
    2\leq k\leq \frac{L-1}{2}, \label{hier3} \\
C_{\frac{L+1}{2},p}&= & C_{\frac{L-1}{2},p}. \label{hier4}
\end{eqnarray}
A striking consequence of momentum conservation is the sum rule
$C_{0,p}+2 \sum_{k=1}^{(L-1)/2} C_{k,p}=0$. Then we expect
that correlations are of the order $O(L^{-1})$. For example, in the
elastic limit $\alpha=1$, their equilibrium value is
$\langle v_{j}v_{j+l}\rangle=-T(L-1)^{-1}$, $\forall l\neq 0$. We take~(\ref{hier}) and~(\ref{hier3}) in the continuum limit,
$x=(k-1)/L$ and $t=p/L^{3}$, and retain only terms up to $O(L^{-1})$,
obtaining
\begin{eqnarray}\label{contcorr}
\frac{dT(t)}{dt}=-\nu \left[ T(t)-L^{-1} \psi(t)\right]+O(L^{-3})\\
\partial_{t}D(x,t)=2\partial_{xx} D(x,t)+O(L^{-2}). \label{contcorr2}
\end{eqnarray}
Here, we have introduced the notations 
$D(x,t)=LC(x,t)$ and $\psi(t)=\lim_{x \rightarrow 0}
D(x,t)$.
Expression (\ref{contcorr}) introduces a correction in the
hydrodynamic average granular temperature, given by the
nearest-neighbour-particle velocity correlation, whereas
(\ref{contcorr2}) is a diffusion equation for spatial
correlations. {Boundary} conditions stem from~(\ref{hier2})
and~(\ref{hier4}), {which give a reflecting boundary at $x=1/2$ and
  the sum rule for momentum conservation,
  $T(t)+2\int_{0}^{1/2}D(x,t)=0$.}  In the long time limit, we obtain
the following scaled stationary solution
\begin{eqnarray}
  \tilde{D}(x)= - A \cos \left[\pi\sqrt{\frac{\nu}{\nu_{c}}} (1 - 2x)
  \right], \quad A
  =\frac{\pi\sqrt{\frac{\nu}{\nu_{c}}}}{\sin\left(\pi\sqrt{\frac{\nu}{\nu_{c}}}\right)},
  \nonumber\\
\label{stcorr1}\end{eqnarray}
where
$\tilde{D}(x)=\lim_{t\rightarrow\infty}D(x,t)/T_{HCS}(t)$. {Note} that
the Fourier transform of $\tilde{D}(x)+\delta(x)$\footnote{The delta
  function is needed to include the case of the autocorrelation $\langle
v_{i}^{2}\rangle$, since $D(0)$ corresponds to $\langle v_{i}v_{i+1}\rangle$,
see the paragraph above \eqref{contcorr}.} takes the form $S(k)=\frac{k^2}{k^2-\nu/2}$ with
$k=2\pi n$ and $n$ is a positive integer, which has been derived for
the correlations of the velocity shear mode in $d>1$ from
{Landau-like} granular fluctuating hydrodynamics, see for
instance~\cite{Noijeernst}. This result reinforces the
  motivation of our 1d model as a simple picture for the shear mode of the
velocities in $d>1$.

\begin{figure} [!h]
 \centering
    \includegraphics[width=0.5\textwidth]{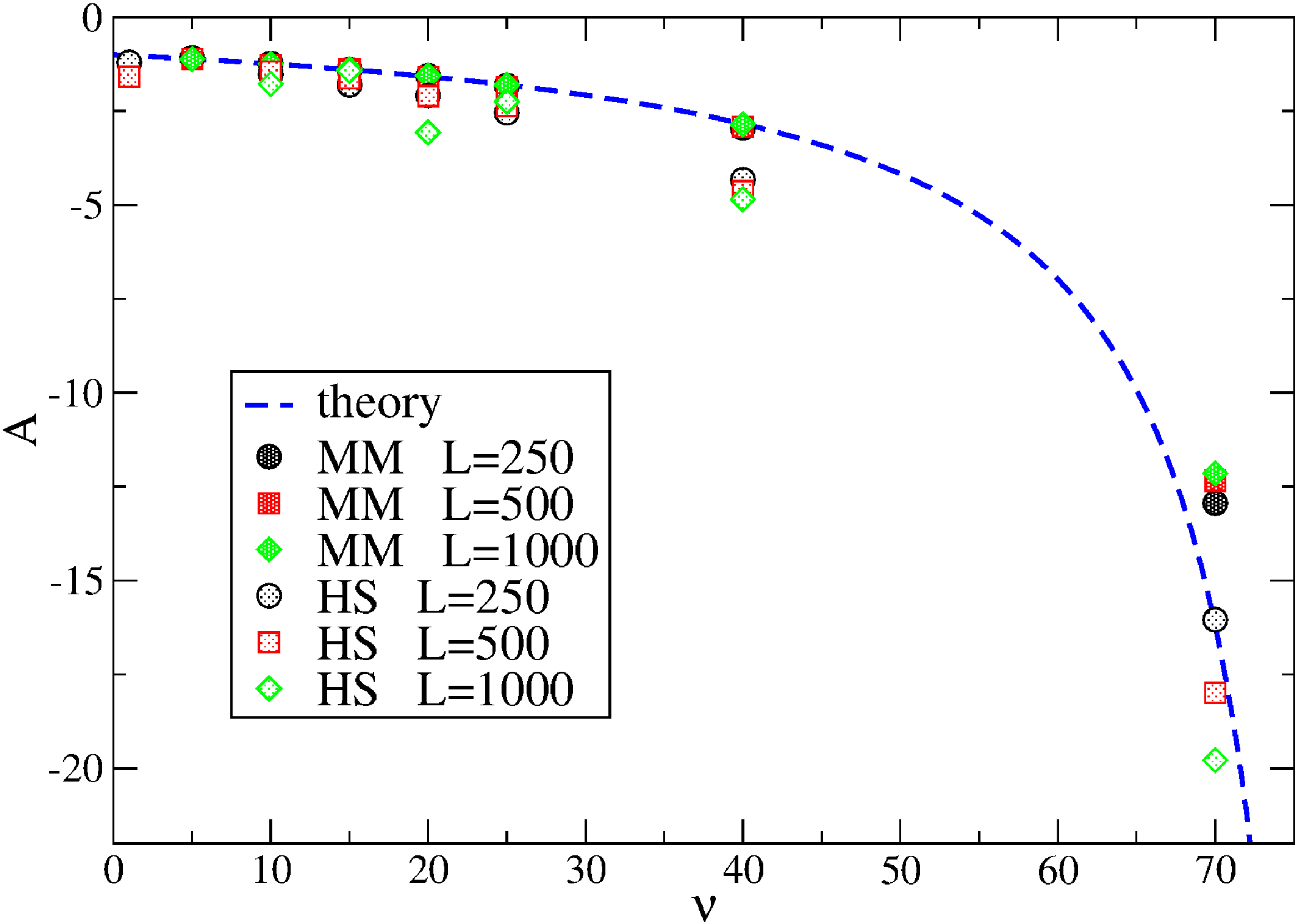}
    \caption{Amplitude $A$ of the rescaled correlation function
      defined in (\ref{stcorr1}) as a function of the dissipation
      rate $\nu$, both for Maxwell molecules (MM)
      and hard spheres (HS). Trajectories start
      from a homogeneous mesoscopic velocity profile
      $u(x,0) \equiv 0$. The theoretical prediction of
      (\ref{stcorr1}) is also shown (line). }
 \label{f:amplitudes}
\end{figure}

In Fig.~\ref{f:amplitudes}, we compare the theoretical prediction in~\eqref{stcorr1} for the MM case with numerical
results. Remarkably, such prediction compares well also with HS
simulations, where the analytical computation appears to be more
challenging. In conclusion, the mechanism that induces spatial
correlations in the system seems to be independent of the particular
interaction model.

\begin{figure}[!h]
    \centering
	\includegraphics[width=0.6\textwidth]{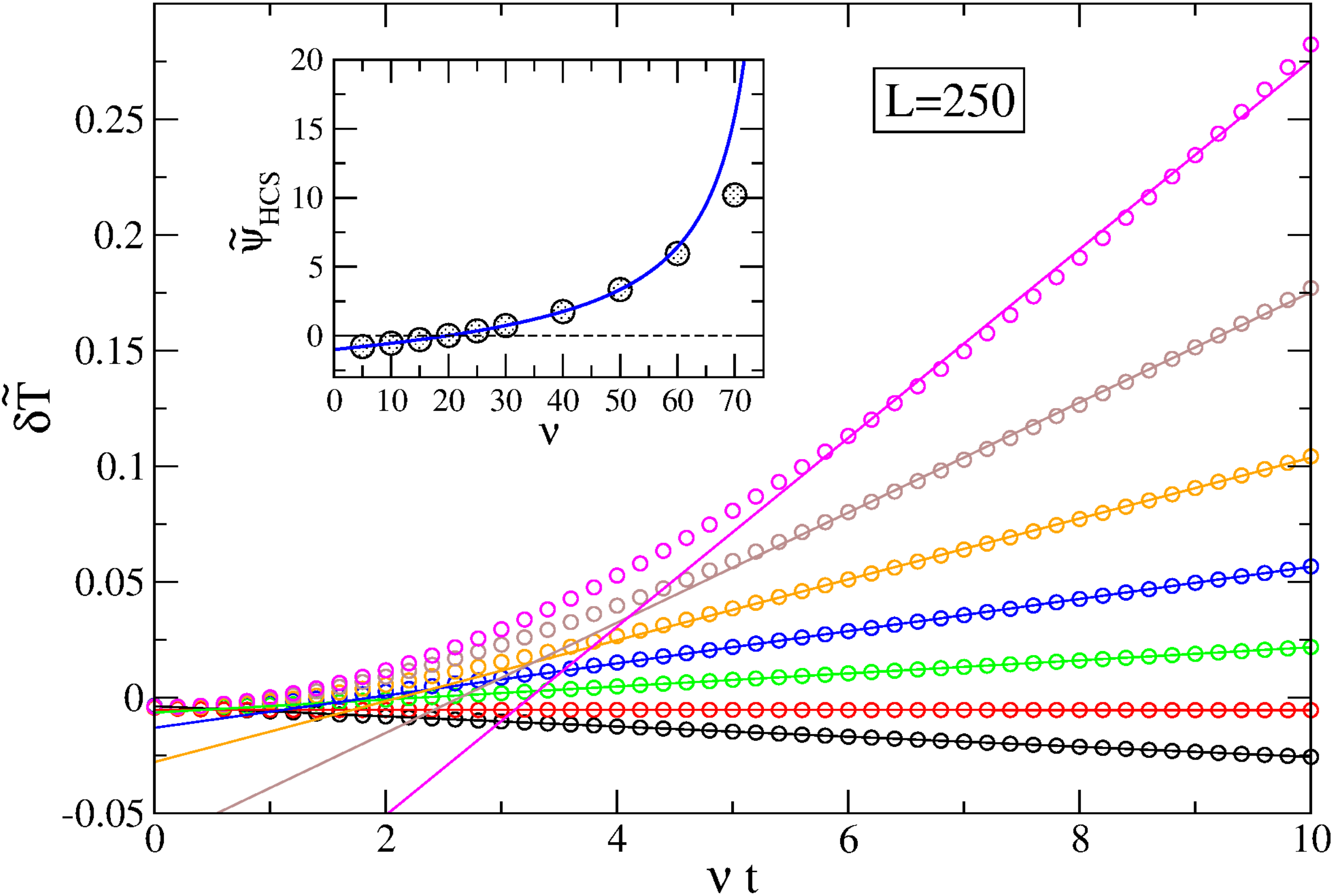}
    \caption{\small  Rescaled temperature $\delta \tilde{T}$ as a
      function of time, for $\nu=10,20,\ldots,70$ (from bottom to
      top). Numerical values (circles) are plotted together with
      linear fits (lines) made on the second half of the trajectory. In the inset, we show the
      comparison between the fitted slopes in the main panel (points)
      and the theoretical ones from~(\ref{Tcorr}) (blue
      line), as a function of $\nu$.  The horizontal black line at
      $\tilde{\psi}_{HCS}=0$ shows the transition at
      $\nu_{\psi} = 2 \pi^2$.}
\label{f:haffviolation}
\end{figure}

Equation~(\ref{contcorr}) suggests that the Haff law
has a finite size correction. We consider a perturbation around
the HCS $T(t)=T_{HCS}(t)+L^{-1} \delta T(t)+... $ and
$D(x,t)=D_{HCS}+L^{-1} \delta D(x,t)+...$. Making use of~(\ref{contcorr}) and defining $\delta \tilde{T} (t) = \delta T (t) / T_{HCS}(t)$, we obtain $\frac{d}{dt} \delta \tilde{T} = \nu \tilde{\psi}_{HCS}$, with
\begin{equation}
\tilde{\psi}_{HCS} =\frac{\psi_{HCS} (t)}{T_{HCS}(t)}= -\pi \sqrt{\frac{\nu}{\nu_{c}}} \cot \left(\pi\sqrt{\frac{\nu}{\nu_{c}}}  \right)  .\label{stcorr2}
\end{equation}
Hence,
 the granular temperature follows
\begin{equation} \label{Tcorr}
T(t) = T_{HCS}(t) \left[ 1 + \frac{1}{L} \tilde{\psi}_{HCS} \nu t + \mathcal{O}(L^{-2}) \right] .
\end{equation}
which is valid for not very long times ($t$ not scaling with
$L$). There is a critical dissipation {value} $\nu_{\psi}=\nu_{c}/4=2
\pi^{2}$ where $\tilde{\psi}_{HCS}$ changes sign, and this determines
a change of the time-derivative of $\delta \tilde{T}$. Thus, at finite
(large) values of $L$, the temperature decays faster or slower than
the Haff law if $\nu<\nu_{\psi}$ or $\nu>\nu_{\psi}$, respectively. In
Fig.~\ref{f:haffviolation}, we compare the predicted Haff law finite
size effect with the simulation results, obtaining excellent
agreement.

\section{Summary.}
In conclusion, we have discussed the rigorous hydrodynamic limit of a
lattice model for granular and active fluids with momentum
conservation and energy dissipation.  Macroscopic equations reproduce
the realistic evolution of the velocity shear mode, which is
diffusive, as well as that of the temperature field, which includes
heat diffusion, inelastic dissipation and viscous heating.  A crucial
phenomenon of inelastic fluids, that is, the shear instability of the
homogeneous cooling state, is recovered.

The model allows us to derive the evolution of $\sim 1/L$ spatial
correlations, which present non-trivial long range extension due to
momentum conservation and alter the temperature decay in an observable
way.  {This opens new interesting paths of investigation, such as
  trying to relate the deviation from the Haff law found here with the
  renormalization of the cooling rate found in systems of smooth HS
  near the shear instability \cite{BDGyM06}.} The appearance of
long-range correlations under non-equilibrium conditions for conserved
fields is a feature well expected on general
grounds~\cite{grinstein,correlations}, but rarely derived in full
analytical detail.

Finally, we stress the importance of considering
finite size effects in granular systems, since large sizes are rarely
realized in experiments. In addition, the HCS is unstable for large
$L$, and therefore finite size corrections cannot be disregarded
by considering an arbitrarily large system. These facts,
  together with the scarcity of studies about finite size effects in
granular matter, makes worthwhile deepening this point in the
next future.

\section{Acknowledgments}
  Antonio Prados acknowledges the support of the Spanish Mi\-nisterio de
  Econom\'\i a y Competitividad through grant FIS2011-24460. We
  would like to thank Carlos A. Plata for useful discussions.

\appendix
\section{Details for the derivation of the hydrodynamic equations}\label{appa}

Average fields (over initial conditions and noise realizations) are
defined as $u_{l,p}\equiv \langle v_{l,p}\rangle$, $E_{l,p} \equiv
\langle v_{l,p}^{2}\rangle$ and $T_{l,p} \equiv E_{l,p} - u_{l,p}^2$.
The microscopic equations for the evolution of averages  at time $p$ at site $l$ are obtained by averaging
equations \eqref{3.4}  and \eqref{3.7} of the main text, obtaining
\begin{eqnarray} \label{microu}
&u_{l,p+1}-u_{l,p}=-\langle j_{l,p} \rangle + \langle j_{l-1,p} \rangle,\\ \label{microE}
&E_{l,p+1}-E_{l,p}=-\langle J_{l,p} \rangle + \langle J_{l-1,p} \rangle + \langle d_{l,p} \rangle.
\end{eqnarray}
Averages of currents and dissipation can be computed assuming the
Local Equilibrium approximation (LEA), which is explicitly stated as
\begin{equation}
P_2(v_l,v_{l+1};p)=\frac{1}{\sqrt{2 \pi T_{l,p}}\sqrt{2 \pi T_{l+1,p}}}e^{-\frac{(v_l-u_{l,p})^2}{2T_{l,p}}-\frac{(v_{l+1}-u_{l+1,p})^2}{2 T_{l+1,p}}}.
\end{equation}
In the Maxwell molecules case ($\beta=0$), where one has
$\langle\delta_{y_p,l}\rangle=1/L$, computations using the LEA
give\footnote{Note that, in the MM case, for obtaining the averages in
  \eqref{av_lea} the LEA is only used to write that
  $\langle v_{l,p}v_{l\pm 1,p}\rangle=u_{l,p}u_{l\pm 1,p}$, that is,
  we assume that velocities at adjacent sites are uncorrelated. This
  hypothesis is somehow similar to the \textit{molecular chaos}
  assumption when writing the Boltzmann equation for a low density
  fluid.}
\begin{subequations}\label{av_lea}
\begin{eqnarray}
 & \langle j_{l,p} \rangle = \frac{1+\alpha}{2L}\left(u_{l,p}-u_{l+1,p}\right),\\
  &\langle J_{l,p} \rangle = \frac{1+\alpha}{2L}\left(T_{l,p}-T_{l+1,p}+u_{l,p}^2-u_{l+1,p}^2\right),\\
 %
&\langle d_{l,p} \rangle  =\frac{\alpha^2-1}{2L} \left[T_{l,p}+\frac{T_{l+1,p}   +T_{l-1,p}}{2}+\left(u_{l,p}-\frac{u_{l+1,p}+u_{l-1,p}}{2}\right)^{2}+\left(\frac{u_{l+1,p}-u_{l-1,p}}{2}\right)^{2}
   \right].
   \label{eq:dlp}
\end{eqnarray}
\end{subequations}

The hydrodynamic limit consists in a change of spatial and time
scales, from $(l,p)$ to $(x,t)$, related by size-dependent factors:
\begin{equation} \label{scal}
x=l/L, \qquad t=p/L^3.
\end{equation}
This implies for a generic function $f_{l,p}$
\begin{eqnarray}
&f_{l+1,p}-f_{l,p}= \frac{1}{L}\frac{\partial}{\partial x} f(x,t) + \mathcal{O}\left(\frac{1}{L^2}\right),\\
&f_{l,p+1}-f_{l,p} = \frac{1}{L^3}\frac{\partial}{\partial t} f(x,t) + \mathcal{O}\left(\frac{1}{L^6}\right),
\end{eqnarray}
which are introduced in~\eqref{microu} and~\eqref{microE} to get
the final continuous equations in $(x,t)$.  Each discrete spatial
derivative introduces a $L^{-1}$ factor in the HL. Then, the
difference between the current terms in the balance equations is of
the order of $L^{-3}$, because the average currents
$\langle j_{l,p}\rangle$ and $\langle J_{l,p}\rangle$ are of the order
of $L^{-2}$. Those terms, therefore, perfectly balance the $1/L^3$
dominant scaling on the left-hand side, i.e. the
time-derivative. Since $\langle d_{l,p}\rangle$ is of the order of
$(1-\alpha^{2})/L$, to match the scaling $1/L^3$ of the other terms,
we define the {\em cooling rate} to be
$\nu=(1-\alpha^2)L^{2}$, which is assumed to be order $1$ when the
limit is taken. This choice automatically implies that when $L$
increases one has that $\alpha$ approaches unity, a
further reason to expect the validity of the LEA.

By retaining only the highest order terms in the equations, we get
expression \eqref{3.16} of the main text. It is interesting
to note that our expansion in terms of $L^{-1}$ is similar in spirit
to the Chapman-Enskog expansion up to Navier-Stokes order,
since we are keeping up to terms of the second order in the gradients
(of the order of $k^{2}$, being $k$ the wave vector, in Fourier
space). From a purely mathematical point of view, \eqref{3.16}
  becomes exact in the limit $L\to\infty$, but
$\nu=(1-\alpha^2)L^{2}$ of the order of unity, as stated in the
previous paragraph. Interestingly, the dissipation field
  $d_{l,p}$ in \eqref{eq:dlp} admits an expansion in even powers of
  the gradients, as is also the case of granular fluids \cite{Brey98,Brill03}. However, in the above limit, the first terms in
  $d_{l,p}$ including the gradients are of the order of $L^{-2}$ as
  compared to the contribution $-\nu T$ at the Navier-Stokes order,
  that is, they would only be considered at the so-called Burnett
  order.

From a physical standpoint, \eqref{3.16} is approximately
valid whenever the terms neglected upon writing it are negligible
against the ones we have kept. Since the correlations
$\langle v_{i}v_{i\pm 1}\rangle$ are expected to be of the order of
$L^{-1}$ as compared to the granular temperature,\footnote{For
  example, in the elastic case, the correlations
  $\langle v_{i}v_{i+k}\rangle$ do not depend on the distance $k$ in
  equilibrium, and therefore
  $\langle v_{i}v_{i+k}\rangle=-T (L-1)^{-1}$, $\forall k\neq 0$. See
  Section~\ref{sec:corr} for more details. } we must impose that
$L\gg 1$ and also $t\ll L$. On the other hand, the term proportional
to the correlations in the evolution equation for the granular
temperature is therefore of the order of $(1-\alpha^{2})L^{-1}$, which
must be negligible against the second spatial derivative terms, of the
order of $L^{-2}$. Then, $(1-\alpha^{2})L\ll 1$ must be further
imposed when the correlations are neglected in \eqref{3.16}. This
condition, although less restrictive that
$1-\alpha^{2}=\mathcal{O}(L^{-2})$, also reinforces the validity of
the LEA. On the other hand, when the correlations are fully taken into
account, as is the case of equations~\eqref{contcorr} and \eqref{contcorr2} of the main paper, the
value of $\alpha$ is not restricted since the only assumption for
writing them is that of homogeneity.

\section{Computation of the correlations of the hydrodynamic noise}\label{appb}

Noises with respect to averages appear in the currents
$j_{l,p} = \widetilde{j}_{l,p} + \xi_{l,p}$,
$J_{l,p}=\tilde{J}_{l,p}+\eta_{l,p}$, and in the dissipation
$d_{l,p}=\tilde{d}_{l,p}+\theta_{l,p}$, with noises $\xi_{l,p}$,
$\eta_{l,p}$ and $\theta_{l,p}$ defined to have zero average. The idea
is that each term $x$ is made of a $\tilde{x}$ part which is an
average over the fast noise (that is, the collisions, which are
counted by the fast stochastic variable $y_{l,p}$), but at fixed
$v_{l,p}$ whose evolution is assumed to be slower than noise.

To obtain the correlations of noise, we exploit a series of
conditions. Explicit calculations are discussed here for the case of
the momentum current noise $\xi_{l,p}$. It is clear that the
definition $\widetilde{j}_{l,p}=(1+\alpha)\Delta_{l,p}/2L$
corresponds to the above prescription for the noise.  First,
it is straightforward that $\langle\xi_{l,p} \xi_{l',p'}\rangle=0$ for
$p\neq p'$, because $y_{p}$ and $y_{p'}$ are independent random
numbers.  Second, we take into account that
$\langle\delta_{y_p,l}\delta_{y_p,l'}\rangle=\delta_{l,l'}\langle
\delta_{y_{p},l}\rangle=\delta_{l,l'}/L,$
and the fact that all the other contributions are of the order of
$L^{-2}$. Thus, for $p=p'$ we have
$ \langle \xi_{l,p}\xi_{l',p} \rangle=
\left(1+\alpha\right)^{2}\left\langle \Delta_{l,p}^{2} \right\rangle
\delta_{l,l'}/4L + O(L^{-2})$.
At this point, the quasi-elasticity of the microscopic dynamics makes
it possible to (i) substitute $(1+\alpha)/2$ by $1$ and (ii) calculate
$\langle \Delta_{l,p}^{2} \rangle$ by using the LEA, to obtain
\begin{equation}\label{4.1.7}
\langle \xi_{l,p} \xi_{l',p'}\rangle \sim \frac{1}{L} \,
2T_{l,p}\,\delta_{l,l'}\, \delta_{p,p'}.
\end{equation}
In the large size system, $j_{l,p}$ scales as $L^{-2}$ (see
\cite{PLyH12a}). Therefore, the mesoscopic noise of the momentum
current is defined as $\xi(x,t)=\lim_{L\to\infty}L^{2} \xi_{l,p}$, and
$j(x,t)=\widetilde{j}(x,t) + \xi(x,t)$, in which, again,
$\widetilde{j}(x,t)=\lim_{L\to\infty}L^{2} \widetilde{j}_{l,p}$. Going
to the continuous limit, remembering~(\ref{scal}), and taking in
account that $\delta_{l,l'}/\Delta x\sim \delta(x-x')$ and
$\delta_{p,p'}/\Delta t \sim \delta(t-t')$ we get the noise amplitude
of the velocity current in the main text. Identical
considerations lead to the amplitude for the energy current
noise.  For the fluctuations of dissipation, the dissipation term is
split again as $d_{l,p}=\tilde{d}_{l,p}+\theta_{l,p},$ with
$\langle d_{l,p}\rangle=\langle\tilde{d}_{l,p}\rangle$. We know from
the dissipation current definition that
$\langle\theta_{l,p}\theta_{l',p'}\rangle=0$ for $p=p'$. Making use of
the LEA and in the large size system $d_{l,p}$ scales as $L^{-3}$ and
it is expected that the noise should have the same
  scaling. Going to the continuous limit and taking in account~\eqref{scal} and~\eqref{4.1.7}, the result in the main text is
recovered.

The cross correlations between different noises are straigthforwardly
obtained, along similar lines, with the result
\begin{eqnarray}\label{4.4.1}
\langle \xi(x,t) \eta(x,t) \rangle=\langle  \eta(x,t) \xi(x,t) \rangle=\frac{4T(x,t)u(x,t)}{L},
\nonumber \\
\langle \xi(x,t) \theta(x,t) \rangle=\langle  \theta(x,t) \xi(x,t) \rangle=0, \nonumber \\
\langle \eta(x,t) \theta(x,t) \rangle=\langle \theta(x,t) \eta(x,t) \rangle=0.
\end{eqnarray}

\section{Numerical comparison for the amplitude of hydrodynamic noises}\label{appc}

\begin{figure} [!h]
 \centering
    \includegraphics[width=0.4\textwidth]{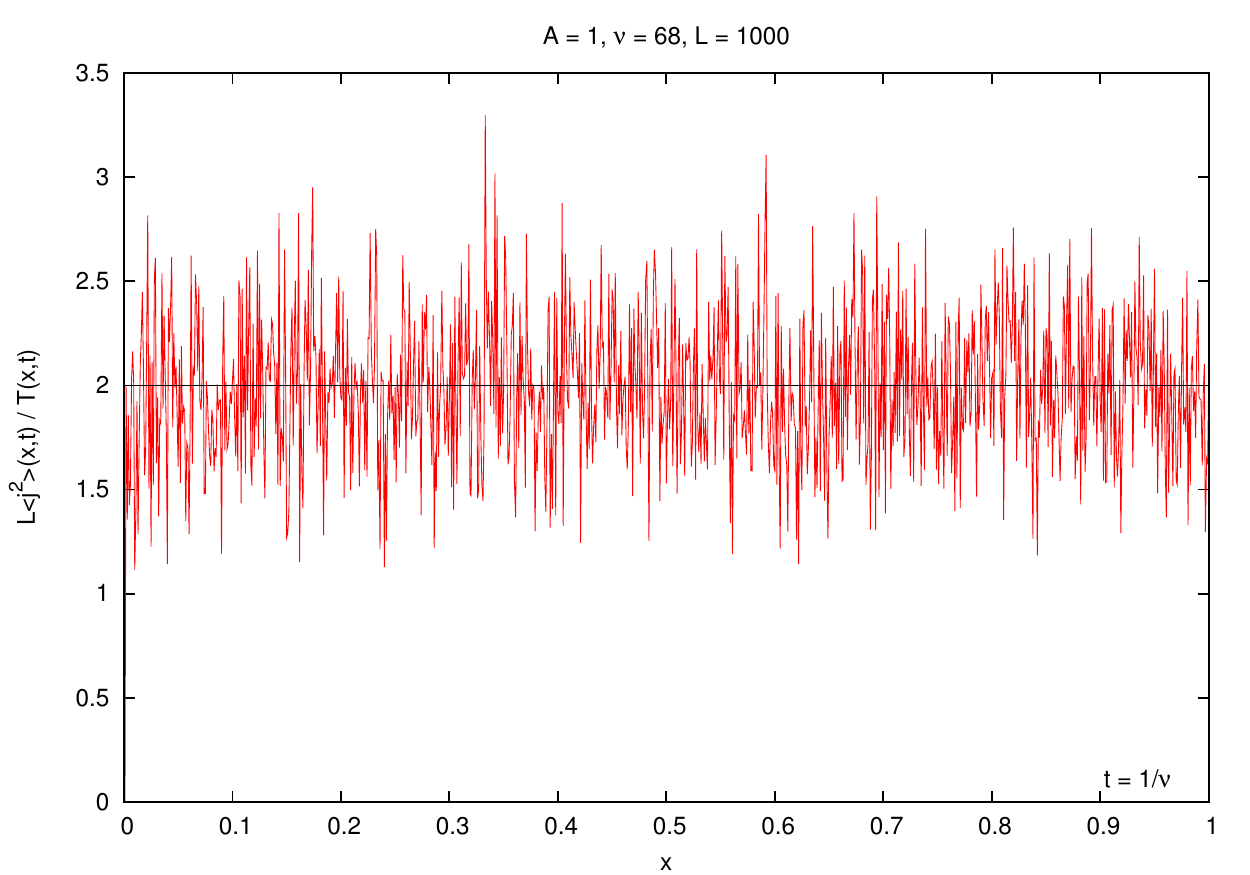}
    \includegraphics[width=0.4\textwidth]{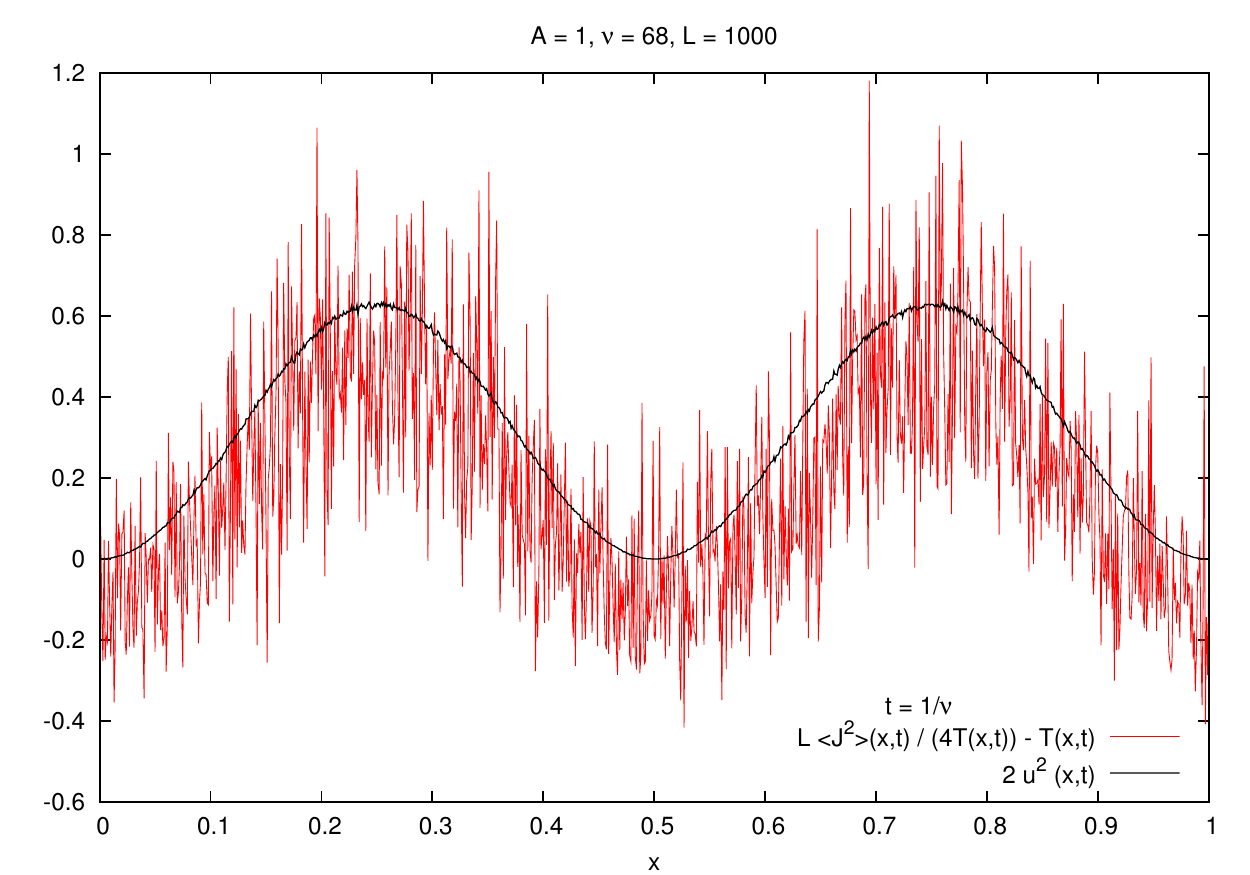}
    \caption{Left panel:  Velocity current fluctuations
      profile rescaled by the local temperature, namely
      $L \langle j^2 (x,t) \rangle / T(x,t)$ as a function of the
      continuum position $x$. The red lines correspond to the
      numerical result, whereas the black line gives the theoretical
      expected value $L \langle j^2 \rangle / T = 2$.  Right panel:
      Energy current fluctuations as a function of $x$. The numerical
      values of $L \langle J^2(x,t) \rangle / 4T(x,t) - T(x,t) $ (in
      red) are compared to the numerical evaluation of the its
      theoretical value in the local equilibrium approximation, $2
      u^2(x,t)$ (in black).  \\
    }
\label{f:noises}
\end{figure}

 A comparison for the amplitudes of noise for the velocity and energy
 currents is shown in Fig.~\ref{f:noises}. A case with the MM
 interaction ($\beta=0$) is considered. The simulations are performed  with periodic boundary conditions, therefore without energy
 injection, and starting with a non-homogeneous initial condition. The initial mesoscopic velocity profile and homogeneous granular temperature are $u(x,0) = u_0 \sin(2 \pi x)$ and
 $T(x,t) \equiv T_0$, respectively, with $u_0 = T_0 = 1$.

\bibliographystyle{ieeetr}
\bibliography{bibliografiav1}

\end{document}